\documentclass[pra,twocolumn,showpacs,amsmath,amssymb]{revtex4}
\usepackage{latexsym}
\usepackage{amsmath, amsthm, amssymb} 

\newcommand{\beq}{\begin{equation}}
\newcommand{\eeq}{\end{equation}}
\newcommand{\ba}{\begin{array}{ccc}}
\newcommand{\ea}{\end{array}}

\newcommand{\bk}{{\bm k}}

\def\bea{\begin{eqnarray}}
\def\eea{\end{eqnarray}}

\usepackage{graphicx}
\usepackage{dcolumn}
\usepackage{bm}
\usepackage{txfonts}
\usepackage{mathrsfs}
\usepackage{feynmf}
\usepackage{comment}
\usepackage{float}

\usepackage{hyperref}
\usepackage{xcolor}
\definecolor{dark-red}{rgb}{0.4,0.15,0.15}
\definecolor{dark-blue}{rgb}{0.15,0.15,0.4}
\definecolor{medium-blue}{rgb}{0,0,0.5}
\hypersetup{
    colorlinks, linkcolor={dark-blue},
    citecolor={dark-blue}, urlcolor={medium-blue}
}

\begin{document}
\title{Soft quantum vibrations of $\mathcal{PT}$-symmetric nonlinear ion chain}


\author{Philipp Strack}
\email{pstrack@physics.harvard.edu}
\affiliation{Department of Physics, Harvard University, Cambridge MA 02138}

\author{Vincenzo Vitelli}
\affiliation{Instituut-Lorentz for Theoretical Physics, Leiden University, Leiden NL 2333 CA, The Netherlands}

\date{\today}

\begin{abstract}
We theoretically study the quantum dynamics of transverse vibrations of a one-dimensional chain 
of trapped ions in harmonic potentials interacting via a Reggeon-type cubic nonlinearity 
that is non-unitary but preserves $\mathcal{P}\mathcal{T}$-symmetry. We propose 
the notion of {\it quantum fragility} for the dissipative structural phase transition that spontaneously breaks 
the $\mathcal{P}\mathcal{T}$-symmetry. In the quantum fragile regime, the nonlinearity dominates the response to mechanical perturbations
and the chain supports neither the ordinary quantum phonons of a Luttinger liquid, nor the supersonic solitons that arise in classical 
fragile critical points in the absence of fluctuations. Quantum fluctuations, approximately captured within a one-loop renormalization group, give rise to mechanical excitations with a strongly momentum-dependent phonon velocity and 
dissipative spectral behavior. Observable signatures of the quantum fragile chain in trapped ion systems are discussed.
\end{abstract}

\pacs{37.10.Ty, 42.65 -k, 43.35.+d, 11.10.Hi}


\maketitle


\section{Introduction}

It has recently become clear that ultracold atom-photon systems offer intriguing opportunities to synthesize a new class of ``soft'' photonic quantum materials such as for example glasses \cite{strack11} and liquid crystals \cite{lechner12} in optical cavities as well as  ``granular'' media made of Rydberg polaritons interacting via hard rod potentials \cite{igor11}. Historically, quantum effects on prototypical soft and granular materials such as sand, polymers, or foams have not been investigated much because they are too large and too heavy and therefore outside the quantum regime. 

However, already on a classical level, the response of certain soft materials is remarkably rich:
granular media at vanishing external pressure or random polymer networks with loose connectivity, for example,
exhibit large deformations in response to applied mechanical perturbations. In both cases, there is a geometrical or topological control parameter that determines the magnitude of the elastic moduli: the average overlap between grains or the mean coordination number of the network. By tuning these parameters one can reach a fragile
\footnote{The notion of fragility that we consider in the present study is not necessarily related to the onset of a plastic response \cite{cates98,fragilebook00}. Instead, we denote a classical or quantum mechanical system as fragile, if a parameter can be tuned such that the quadratic part of its Hamiltonian becomes sub-dominant so that its response to mechanical perturbations is intrinsically non-linear.}
, mechanical state characterized by a vanishing linear response \cite{Epitome,MvHecke,XuEPL}, often termed sonic vacuum since the linear speed of sound vanishes \cite{nesterenko84,Nesterenko_Book,Gomez_2012,gomez12}. In these fragile states, even the tiniest strains propagate as supersonic solitons (or shocks) rather than ordinary phonons. 

In this paper, we propose a new quantum soft matter phase in the fragile regime. Extending the line 
of work on quantum simulation with phonons \cite{porras04} and structural, ``Zigzag'' quantum phase transitions in ion chains \cite{retzker08,shimshoni11}, we here consider the quantum phase transition from a one-dimensional solid of ions to a dissipative state with broken $\mathcal{P}\mathcal{T}$-symmetry induced by a Reggeon-type cubic nonlinearity coupling neighboring ions. As in the known soft counterparts, the  mechanical response (the quadratic term in their Hamiltonian) of our model can become vanishingly small such that the response is predominantly nonlinear; on top of that, the mechanical properties are strongly affected by zero point fluctuations leading to a quantum phase transition in the Reggeon universality class \cite{cardy77,moshe77,moshe78}.

\begin{figure}[t]
\vspace{5mm}
\includegraphics*[width=75mm,angle=0]{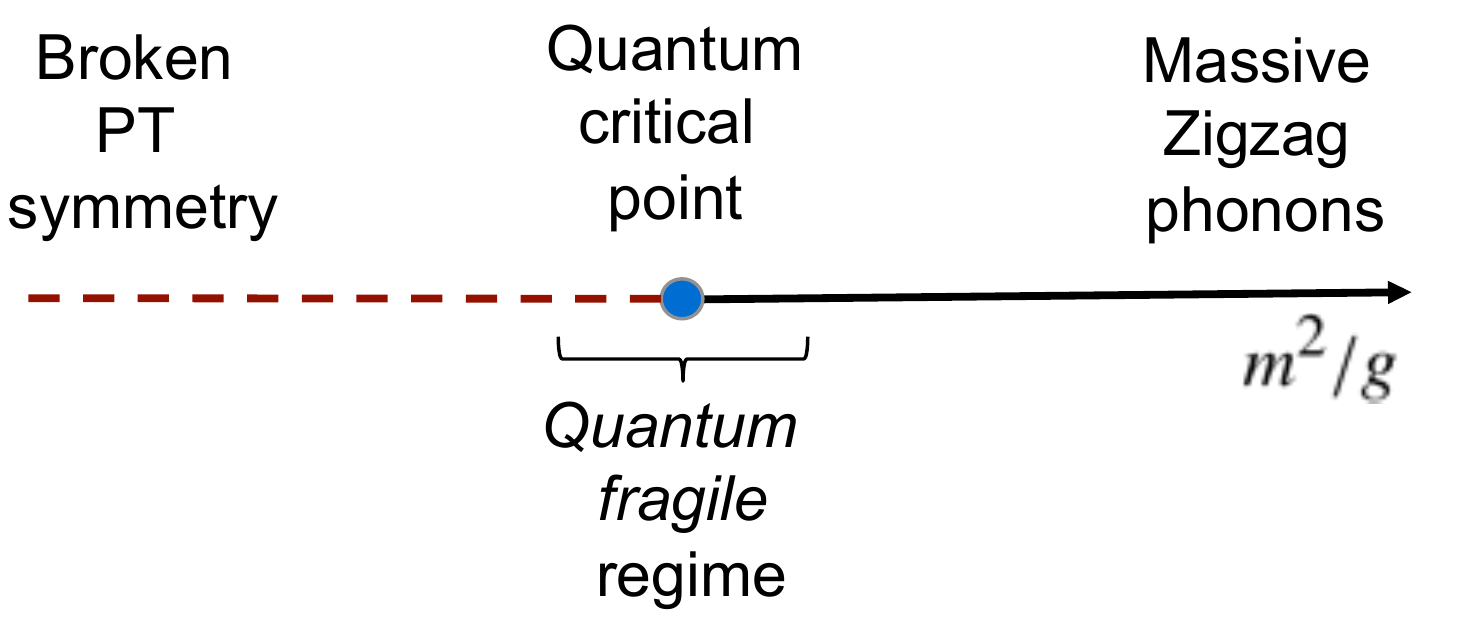}
\caption{(Color online) Zero-temperatures phase diagram of Reggeon quantum chain. For large harmonic potential (to the right 
of the plot), the chain exhibits a massive Zigzag mode and exponentially decaying correlations 
$\langle \phi_\ell \phi_m\rangle \sim e^{- |x_i - x_j| m}$. In the regime around the quantum critical point (blue dot), 
the chain is quantum fragile with long-ranged, power-law correlations among transverse vibrations 
$\langle \phi_\ell \phi_m \rangle \sim \frac{1}{ |x_i - x_j|^{2-\eta^\ast_A}}$ with $\eta^\ast_A$ computed 
in Eq.~(\ref{eq:eta_A}).
}
\label{fig:phase_diag}
\end{figure}

The main physical result of our analysis is that the basic excitations at the quantum 
critical point of the Reggeon ion chain (illustrated in Fig.~\ref{fig:phase_diag})
are neither ordinary phonons, that cannot exist because of the vanishing (transverse) rigidity, nor the strongly non-linear solitons that characterize similar critical points in the absence of fluctuations. Instead, quantum fluctuations give rise to mechanical excitations with a non-linear dispersion relation. This can be traced to the emergence of a length-scale dependent 
propagation velocity of transverse waves (or equivalently elastic modulus). 

Using the renormalization group to one-loop order, we can actually flow into the highly nonlinear regime close to the quantum critical point, where violent fluctuations of the ion displacements lead to Reggeon-type quantum criticality that is neither Luttinger liquid-like nor part of the $1\!+\!1$-dimensional Ising universality class of the Zigzag transition \cite{shimshoni11}. It is well known that the presence of a continuous symmetry in combination with unitarity inevitably leads to Luttinger liquid behavior of conventional gapless quantum liquids in one dimension \cite{haldane81,giamarchi04}. Both of these constraints are relaxed in the highly non-linear regime of the non-unitary quantum chain. This is the reason why the imaginary cubic nonlinearity acts as a relevant perturbation to the Luttinger liquid behavior and leads to an unusual power-law scaling of correlators. 

At the quantum critical point, when the chain becomes fragile, the low-energy, low-momentum, mechanical excitations 
disperse at low 
as 
\begin{align}
\omega \sim k^{z}\;\,\,\text{with}\;\;z=0.35 
\end{align}
signaling a strongly momentum-dependent propagation velocity (recall that for conventional phonons $\omega \sim v_s | \bk | $). The associated spectral function becomes dissipative and displays a quantum critical continuum that may associated with the spontaneous breakdown of the $\mathcal{P}\mathcal{T}$-symmetry. We propose to associate this new quantum critical point with the notion of quantum fragility generalizing the physics known from soft materials to the quantum regime. 
In the phase with broken $\mathcal{P}\mathcal{T}$ 
symmetry, some of the energy eigenvalues \cite{bender12}, 
as well as potentially some of the mechanical response functions, 
can become complex-valued. As in Ref.~\onlinecite{shimshoni11}, we expect the phase transition to be observable via measurements of the dynamic structure factor via light scattering.

Our main motivation to study the relatively exotic Reggeon ion chain is that it displays
appealing, critical features that have not been discussed previously in the trapped ion quantum simulation community \cite{roos08,prutti11,brown11,cirac12,schneider12}. Realization of the Reggeon ion chain 
or related models such as the Fermi-Pasta-Ulam problem \cite{gallavotti08} in a regime where both, the strongly non-linear mechanical response as well as quantum fluctuations are important, could open up the new field of studying fragile materials in a quantum regime. Originally, 
Fermi, Pasta, and Ulam studied N classical oscillators with coordinates $x_\ell$ and momenta 
$p_\ell$
\begin{align}
\sum_{\ell=1}^{N-1} 
\frac{p_\ell^2}{2\mu}
+
\sum_{\ell=1}^N \frac{1}{2} \left(x_{\ell+1}-x_\ell\right)^2
+\frac{\alpha}{3}\left(x_{\ell +1} - x_\ell\right)^3
+\frac{\beta}{4}\left(x_{\ell + 1} - x_\ell\right)^4\;
\end{align}
with the aim to relate the ergodic hypothesis in phase 
space to the presence of nonlinearities $\alpha$, $\beta$ \cite{gallavotti08}.
We will below study a {\it quantized} version with $\beta=0$, $\alpha$ complex,
and an additional pinning potential.

The connection of our model to trapped ion chains relies on works on structural 
phase transitions of strings of charged particles in harmonic traps by Morigi, Fishman and others
\cite{morigi04,fishman08,retzker08,shimshoni11}. The structural transition is 
driven by {\it transverse vibrations} $\hat{x}$ of the ions about their equilibrium positions. The interesting 
interactions/non-linearities are generated by a combination of movement of the ions in the {\it axial} directions of the harmonic potential and experiencing mutual Coulomb interaction proportional to the square of their charge $Q^2$. The main extension of our model is that rather than the quartic, local 
interaction $\hat{x}^4$ interaction previously considered \cite{retzker08,shimshoni11}, we will 
explore the more exotic possibility of an imaginary, cubic nearest-neighbor interaction 
$i g \left(\hat{x}_\ell - \hat{x}_{\ell +1}\right)^3$. In terms of experimental 
scales and parameters, we will be operating in the same regime as Refs.~\onlinecite{retzker08,shimshoni11}.

The rest of the paper is structured as follows. In Sec.~\ref{sec:model}, we present the Hamiltonian of our model and 
its equivalent quantum field theory formulation. In Subsec.~\ref{subsec:method}, we explain our 
renormalization group approach and show how the flow equations yield an interacting fixed point describing the phase transition and its critical exponents.
The new physics associated with this fixed point and its observable signatures are presented in 
Sec.~\ref{sec:observables}.
In Sec.~\ref{sec:conclu}, we close with a short conclusion.

\medbreak

\section{Model}
\label{sec:model}

Our starting point are $N$ ions of mass $\mu$ and charge $Q$ confined in a two-dimensional plane 
by an anisotropic harmonic potential mutually interacting via a static Coulomb interaction 
\cite{retzker08,shimshoni11}. The Hamiltonian is
\begin{align}
\hat{H} = \sum_{\ell = 1}^N \frac{\hat{p}^2}{2\mu} + \frac{\mu \nu^2_t}{2} \hat{x}^2_\ell+
\frac{Q^2}{2} \sum_{\ell \neq m} \frac{1}{\sqrt{\left(x_\ell - x_m\right)^2 + \left(z_\ell - z_m\right)^2}}
\end{align}
with motion in the $z$-direction is frozen out the by a tight confining potential. $\hat{p}_\ell$ is the conjugate momentum operator fulfilling the discrete commutation relation: $\left[\hat{p}_{\ell}, \hat{x}_{\ell'}\right]=-i\delta_{\ell \ell'}$ ($\hbar = 1$ Along the $z$-axis the 
equilibrium positions of the particles are equidistantly aligned with (lattice constant) $a$. One can now perform a gradient expansion and obtain an effective model for the transverse vibrations 
in $x$-direction; the coefficients of this expansion depend on the trapping potential and the Coulomb interaction \cite{shimshoni11}.
We will continue with the effectively one-dimensional Hamiltonian for the transverse displacements 
from the equilibrium position
\begin{align}
\hat{H} = \sum_{\ell=1}^N  \frac{\hat{p}_{\ell}^2 }{2 \mu}
+ 
\frac{m^2}{2} \, \hat{x}_{\ell}^2
+
\frac{A}{2} \left(\hat{x}_\ell - \hat{x}_{\ell + 1}\right)^2
+ 
i \frac{g}{36} \left(\hat{x}_{\ell} - \hat{x}_{\ell + 1}\right)^3\;.
\label{eq:hamilton}
\end{align}
where the coefficients of the harmonic restoring force $A$ and the quadratic part of the 
effective pinning potential $m^2$ in $x$-direction can be related to microscopic parameters \cite{shimshoni11}
\begin{align}
A&=\frac{Q^2}{a} \text{ln} 2
\nonumber\\
m^2&=-\mu\left(\nu_c^2 - \nu_t^2\right) a^2
\nonumber\\
\nu_c&=\sqrt{\frac{4 Q^2}{ma^3}}\sqrt{C_3}
\label{eq:paras}
\end{align}
with $C_\alpha = \sum_{\ell \geq 1} \frac{1}{\left(2 \ell + 1\right)^\alpha}$. As announced above, 
we here consider as an input the cubic nonlinearity $ig$ and explore its qualitatively new consequences. 
In the future, it will be interesting to add higher-power interaction terms and study their interplay.

\begin{figure}[]
\vspace{8mm}
\includegraphics*[width=85mm,angle=0]{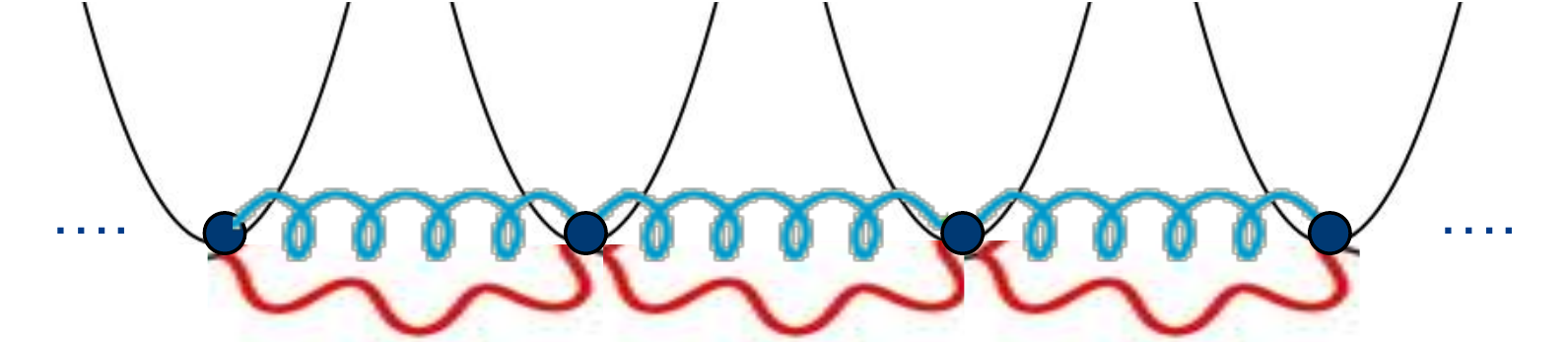}\\[8mm]
\caption{(Color online) Sketch of the model Eq.~(\ref{eq:hamilton}) for the transverse vibrations 
of a chain of trapped ions in a harmonic 
potential connected by nonlinear ``springs''. Blue, straight springs denote the effectively one-dimensional
harmonic restoring forces arising due to a combination of trapping potential and Coulomb interaction between the 
charged ions. Red, curved springs denote the (imaginary) cubic nonlinearity between neighboring 
ions. The main physical effect of the blue, harmonic force is to generate phonons as in the 
textbook harmonic chain. The red, nonlinear forces tend to make the chain fragile and eventually lead 
to the dissipative structural phase transition of the Zigzag mode. We have chosen the spring representation 
to highlight cross-connections to (floppy) polymer networks in soft matter physics.}
\label{fig:chain}
\end{figure}

The Hamiltonian in Eq.~(\ref{eq:hamilton}) is not Hermitian -- it is not invariant under combined matrix transposition and complex conjugation. The transversely projected Hamiltonian is sketched in Fig.~\ref{fig:chain}. In addition to translational invariance in space $\ell\rightarrow \ell + n$ with $n$ integer and time, Eq.~(\ref{eq:hamilton}) is invariant under a simultaneous parity transformation ($\mathcal{P}$:
$\hat{x}\rightarrow - \hat{x}$) and complex conjugation or time-reversal ($\mathcal{T}$).

Three noteworthy features of Eq.~(\ref{eq:hamilton}) are:  

(i) Reggeon-type, non-unitary field theories with cubic nonlinearity have 
a successful history of describing geometric phenomena such as directed percolation \cite{moshe78,hinrichsen00,saleur96}. 

(ii) The cubic nonlinearity $ig$ is chosen proportional to the strain similar to the granular chain \cite{nesterenko84} but with an 
integer-valued slightly above the Hertz law for spheres where the interaction scales as 
\begin{align}
\sim\left(x_\ell - x_{\ell + 1}\right)^\alpha
\label{eq:hertz}
\end{align}
with $\alpha=5/2$.\footnote{Note that unlike the grains, the interaction considered in Eq.~(\ref{eq:hamilton}) has both an attractive as well as repulsive component.}

(iii) The initially non-zero mass from the pinning potential explicitly breaks the 
continuous shift invariance $\hat{x}_{\ell}\rightarrow \hat{x}_{\ell}+\delta$ thereby avoiding the Luttinger liquid fixed point 
\footnote{We thank S. Sachdev for pointing this out.} of gapless one dimensional quantum systems\cite{haldane81}

The cubic nonlinearity preserves $\mathcal{P}\mathcal{T}$-symmetry and also simplifies the mathematical analysis, 
for example by giving rise to anomalous dimensions already at the one-loop level. Bender and Boettcher \cite{bender98,zinn10} demonstrated that certain, non-Hermitian, imaginary Hamiltonians, like the one in Eq.~(\ref{eq:hamilton}), can still describe sensible quantum mechanical ground states with real and positive spectra provided they fulfill a combined parity and time-reversal, $\mathcal{P}\mathcal{T}$-symmetry. Such $\mathcal{P}\mathcal{T}$-symmetric systems have since been realized experimentally in optical waveguides \cite{guo09} and 
it seems possible to engineer periodically structured nonlinearities as well \cite{miri12}. 
Related chains of oscillators with Hermitian nonlinearities such as the Fermi-Pasta-Ulam problem \cite{gallavotti08}, quantum Frenkel-Kontorova chains \cite{zhirov03,zhirov11}, chains with disorder \cite{dyson53}, or chains under applied forces \cite{coppersmith88} can also be used as model systems to access the strongly non-linear regime considered 
in the present study.


In a potential experiment, one would start with linear chain (sketched in Fig.~\ref{fig:chain}), 
and then tune through the critical point for example by chaning the transverse frequency 
$\nu_t$ such that the mass term in Eq.~(\ref{eq:paras}) becomes small. The structural re-arrangement 
of the chain and long-ranged, spatial correlations should be visible in 
fluorescence spectroscopy. Experimental challenges are: (i) frequency resolution 
which should be better than kHz to several Hz for depending on the mass of the ions used 
to detect the critical dynamics \cite{shimshoni11}, (ii) system size and homogeneity, we expect similar 
number of ions (of order $\sim$ 10) to be necessary in line with recent 
experiments on a trapped ion realization of the quantum Ising model \cite{islam11}, 
and (iii) achieving sufficiently small temperatures of the order of $\mu$K to $mK$ 
depending on the mass of the precise ions used. Precise values of trapping frequencies 
will also depend on the detailed experimental apparatus and ions used.

Therefore, we conclude 
that the quantum dynamics of the nonlinear ion chain Eq.~(\ref{eq:hamilton}) 
seems experimentally feasible but will face essentially the same challenges than 
other experimental quantum simulations with trapped ions \cite{schneider12}.

\subsection{Quantum field theory}
For vanishing strength of the harmonic potential $m^2\rightarrow 0$ in 
Eq.~(\ref{eq:hamilton}), the transverse vibrations $\hat{x}_\ell$ develop long-ranged correlations 
ranging over many ions in the chain. The Zig-zag mode becomes soft and 
the chain susceptible to ``mechanical'' perturbations. In this critical 
regime, quantum mechanical perturbation theory breaks down due to the smallness of 
energy denominators. Additionally, it is not a simple matter to 
quantize the problem of classical structural transitions in the critical regime 
because normal modes may not be well defined. In the sonic 
vacuum, for example, there is no sound mode and one cannot quantize phonons in the canonical 
fashion by promoting the normal modes to operators fulfilling commutation relations.

We here capture quantum fluctuations of the ion's transverse position, 
(the ions are expected to undergo zero-point motion at sufficiently low temperature), 
by associating the quantized position coordinate $\hat{x}_\ell$ with a {\it pseudo-scalar 
quantum field }
\begin{align}
\hat{x}_\ell \leftrightarrow \phi(\tau)_\ell
\end{align}
with domain, $-\infty < \phi_\ell (\tau) < + \infty$, living on a one-dimensional lattice with $N$ sites. 
$\tau$ is the imaginary time coordinate ranging from $0\leq\tau<1/T \rightarrow \infty$ for zero temperature $T=0$. 
The fields fulfill periodic boundary conditions $\phi_\ell (0) = \phi_\ell (1/T)$ \cite{negele88}. Fluctuations 
of $\phi$ are integrated over in the functional integral representation of the partition function
\begin{align}
Z = \int \mathcal{D}\{ \phi \}\,e^{-S[\phi]}
\label{eq:Z}
\end{align}
with the lattice action given by
\begin{align}
S[\phi]=& \int_0^{1/T}\! \!d \tau \sum_{\ell = 1}^N
\Bigg[
 \frac{1}{2} \Big( \partial_\tau \phi_\ell(\tau)\Big)^2 + \frac{m^2}{2} \Big(\phi_\ell(\tau)\Big)^2
 \nonumber\\
&+ \frac{A}{2} \Big(\phi_\ell(\tau)-\phi_{\ell + 1}(\tau)\Big)^2
+ i \frac{ g }{36} \Big(\phi_\ell(\tau)-\phi_{\ell + 1}(\tau)\Big)^3
\Bigg]\;.
\label{eq:lattice}
\end{align}
Following Ref.~\onlinecite{bender12}, we assume $\phi$ to transform as a pseudoscalar, that is, $\phi$ changes sign under space reflection $\mathcal{P}$;  then the interaction remains $\mathcal{PT}$-invariant since $i$ changes sign under $\mathcal{T}$. We assume the cubic interaction be local in time; retardation effects from the photon-mediated 
Coulomb interaction between neighboring ions are subdominant here.

Compared to the Hamiltonian formulation Eq.~(\ref{eq:hamilton}), the 
field theory Eqs.~(\ref{eq:Z},\ref{eq:lattice}) is more convenient to compute dynamical 
properties due to the (imaginary) time dependence of the fields. Moreover, we can now use the Wilsonian 
renormalization group (RG) to compute correlation functions in the critical (fragile) regime.

\section{One-loop renormalization group} 
\label{sec:RG}
%

In order to perform our RG analysis we write the Lagrangian in a Fourier representation using
%
$\phi_\ell(\tau)=\sum_{k} \phi_k(\tau) e^{i k x_\ell}\;,$
%
and analogously for frequencies $\omega$. The lattice action in the thermodynamic limit is obtained as:
%
\begin{align}
\Gamma_{\Lambda_0}[\phi]=&
\int_{\omega,k}
\phi_{-k}(-\omega)
\Bigg(
 \frac{Z_{\Lambda_0}}{2}\omega^2 + A_{\Lambda_0} \left(1-\cos[k]\right)
+
\frac{m_{\Lambda_0}^2}{2}
\Bigg)
 \phi_{k}(\omega)
 \nonumber\\
 &
 -
 \frac{g_{\Lambda_0}}{6} 
 \int_{\tau,k_2,k_3}
 \sin \left[k_2 + k_3\right]
 \phi_{-(k_2+k_3)}(\tau) \phi_{k_2}(\tau)\phi_{k_3}(\tau)\;,
 \nonumber\\
\label{eq:fourier}
 \end{align}
%
where we have abbreviated $\int_{\omega,k}=\int^{\infty}_{-\infty} \frac{d \omega} {2\pi}\int^{\pi}_{-\pi} \frac{ d k}{2\pi}$ 
and $ \int_{\tau,k_2,k_3}=  \int^{\infty}_0 d \tau \int^{\pi}_{-\pi} \frac{d k_2}{2\pi} \int_{-\pi}^{\pi} \frac{d k_3}{2\pi} 
$.
We have introduced already here the cutoff scale $\Lambda \in \{\Lambda_0,0\}$, where 
$\Lambda_0$ is the ``ultraviolet'' momentum scale of the order of the inverse lattice spacing at which the renormalization
parameters $Z_{\Lambda_0}$, $A_{\Lambda_0}$, $m^2_{\Lambda_0}$ and $g_{\Lambda_0}$ take their initial values. 
The endpoint of the RG flow is at $\Lambda\rightarrow 0$ and 
quantum fluctuations on all scales have been integrated into the effective action $\Gamma_{\Lambda}[\phi]$. 
We exclude large momentum transfers from the analysis and approximate the trigonometric functions in Eq.~(\ref{eq:fourier}) 
by their leading polynomials
%
$1-\cos[k]\rightarrow \frac{k^2}{2} + ...$ and 
$\sin[k_2 + k_3 ]\rightarrow k_2 + k_3 + ...$.
%
Now, the cubic vertex becomes proportional to the total momentum that flows through it. 

Note that the propagator in the quadratic term in the first line of Eq.~(\ref{eq:fourier}) has a massive phonon dispersion, leading to spectral weight at  
\begin{align}
\omega_{\text{massive}} = \sqrt{\frac{A k^2 + m^2}{Z}}\;,
\label{eq:massive}
\end{align}
with $Z$, $A$, and $m^2$ attaining finite values. Throughout the massive phase, the position of the peak will receive finite renormalizations but otherwise the theory will still support well-defined quasi-particles.
When approaching the critical point, for vanishing mass $m^2\rightarrow 0$, one immediately notices that a loop expansion of Eq.~(\ref{eq:fourier}) will suffer from logarithmic and power-law infrared singularities. We will see below that the RG flow to low frequencies and momenta in the presence of interactions gives rise 
to fractional gradients in space and time characterized by two different anomalous dimensions. 

\subsection{RG Scheme}
\label{subsec:method}

\begin{figure}[t]
\includegraphics*[width=75mm,angle=0]{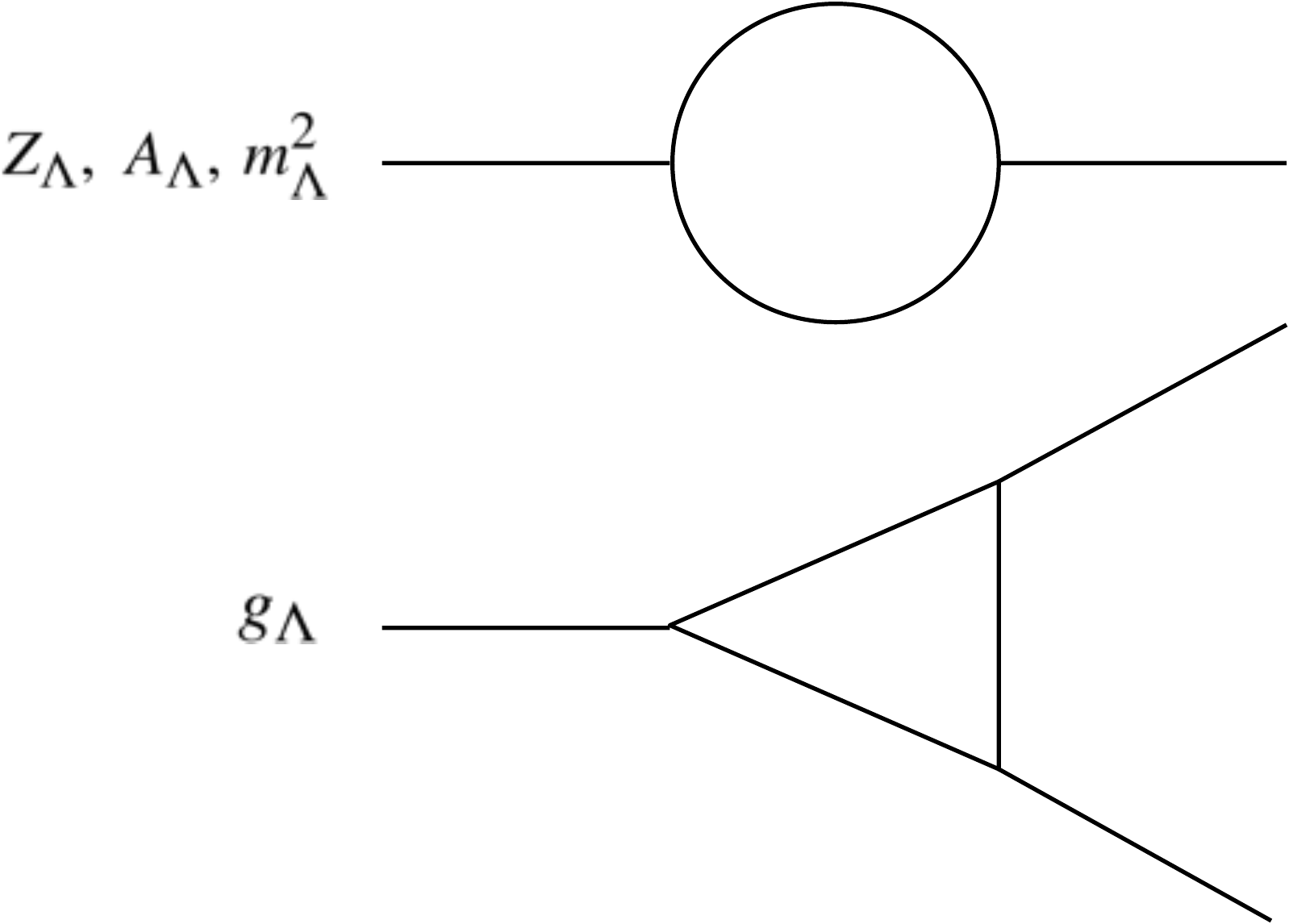}
\caption{Feynman diagrams for the flow equations to capture 
quantum fluctuations of the transverse vibrations in the chain. The straight line 
denotes the Zigzag propagator and the crossing of three lines the 
cubic interaction vertex.}
\label{fig:diags}
\end{figure}

To analyze the physics of these singularities, we employ the (formally exact) RG flow equation for the effective action $\Gamma_\Lambda\left[\psi\right]$, 
the generating functional for one-particle irreducible correlation functions in the form derived by Wetterich \cite{berges02,strack09}. We mainly use this method due to its simplicity and versatility to tailor it to the problem at hand. In principle, similar results should be obtainable within an $\epsilon$-expansion or a field-theoretic RG. We add a regulator $R_{\Lambda}$ (specified below) to the quadratic part of our bare action Eq.~(\ref{eq:fourier}) that introduces a cutoff dependence into the effective action so that 
$\Gamma_{\Lambda}[\phi]$ smoothly interpolates between the bare action, 
Eq.~(\ref{eq:fourier}), at the ultraviolet scale 
$\Gamma_{\Lambda=\Lambda_{0}} \left[\phi \right] =\Gamma_{\Lambda_{0}}[\phi]$ 
and the fully 
renormalized effective action in the limit 
of vanishing cutoff: $\lim_{\Lambda\rightarrow 0} \Gamma_{\Lambda}\left[\phi\right] = \Gamma\left[\phi\right]$. 
The Wetterich equation 
\begin{align}
\partial_{\Lambda} \Gamma_\Lambda[\phi] = \frac{1}{2} \text{Tr} 
\Big[
\frac{\partial_{\Lambda} R_\Lambda}{
\Gamma^{(2)}_\Lambda[\phi]  + R_{\Lambda}}
\Big]
\label{eq:wetterich}
\end{align}
has a one-loop structure and in a vertex expansion the $\beta$-functions for the $n$-point correlators 
are determined by (cutoff derivatives of) one-particle irreducible one-loop diagrams with fully dressed propagators 
and vertices. The Tr is here just a frequency and momentum integration and $\Gamma^{(2)}_\Lambda[\phi]=
\frac{\partial^{2}\Gamma_{\Lambda}[\phi]}{\partial \phi ^2}$ is the 
second functional derivative with respect to the fields $\phi$.
In the course of the RG flow, new terms, not present in the bare action, will be generated. 
The objective of the present 
paper is to choose a simple truncation that reveals the qualitatively new physical features arising from the interplay of quantum fluctuations and the imaginary nonlinearity. We 
therefore focus on the flow of the scale-dependent 
phonon frequency renormalization factor $Z_\Lambda$, the momentum renormalization factor $A_{\Lambda}$, the mass $m^2_{\Lambda}$ and the 
non-linear coupling $g_{\Lambda}$. No local cubic interaction $ig_{\text{loc}}\phi_{\ell}^3$ is generated at one-loop. This is because the k-integration in the triangle diagram (see Fig.~\ref{fig:diags}) picks up a factor $k^3$ and vanishes by symmetry.

As an infrared regulator we use a mass-like cutoff
%
$R_\Lambda = A_{\Lambda} \Lambda^2$ so that
$\partial_\Lambda R_{\Lambda} = 2 A_{\Lambda}\Lambda$
%
where, as customary, we drop the (higher-order) term $\sim\partial_{\Lambda}{A}_{\Lambda}$ in the scale-derivative of the cutoff. 
Now, the infrared singularities at low frequencies and momenta of the phonon propagator
\begin{align}
G_{\Lambda}(\omega,k)
=
\left[\Gamma^{(2)}_\Lambda[\psi]  + R_{\Lambda}\right]^{-1}
=\frac{1}{Z_{\Lambda}\omega^2 + A_{\Lambda} k^2 + m^2_{\Lambda}+R_\Lambda}\;
\end{align}
are regulated  without by-hand changing the ``relativistic'' scaling of the bare phonon dispersion $\omega \sim k$. 
We now plug in Eq.~(\ref{eq:fourier}) supplemented 
by the mass and regulator terms into Eq.~(\ref{eq:wetterich}). By comparing coefficients of the fields in a vertex expansion, we obtain the flow equations 
for $Z_{\Lambda}$, $A_{\Lambda}$, $m^2_{\Lambda}$, and $g_{\Lambda}$. The corresponding diagrams are shown 
in Fig.~\ref{fig:diags}.

Upon using rescaled variables, the flow equations can be brought into a simple form that facilitates comparison to the well known Wilson-Fisher 
fixed point of the $\phi^4$ theory below its upper critical dimension (see for example Ref.~\onlinecite{chaikin_book}). The anomalous dimensions absorb singular renormalizations to the 
frequency ($Z$) and momentum ($A$) factor of the phonon propagator, respectively:
\begin{align}
\eta^Z_\Lambda=-\frac{\Lambda}{Z_{\Lambda}} \partial_{\Lambda}Z_{\Lambda}
\;,\;\;\;\;\;\;
\eta^A_\Lambda=-\frac{\Lambda}{A_{\Lambda}} \partial_{\Lambda}A_{\Lambda}\;.
\end{align}
Upon rescaling the mass and the cubic interaction as
\begin{align}
\tilde{m}_{\Lambda}^2 = \frac{m_{\Lambda}^2}{A_{\Lambda}\Lambda^2}
\;,\;\;\;\;\;\;
\tilde{g}_{\Lambda}=\frac{g_{\Lambda}}{\Lambda  A_{\Lambda}^{5/4} Z_{\Lambda}^{1/4}}\;,
\end{align}
we obtain their $\beta$-functions: 
\begin{align}
\Lambda \partial_{\Lambda} \tilde{m}^2_{\Lambda}
&=
\left(-2 + \eta^A_{\Lambda}  \right) \tilde{m}^2_{\Lambda}-\frac{1}{8\pi} \tilde{g}^2_{\Lambda} \frac{1}{1+\tilde{m}^2_{\Lambda}} 
\label{eq:mass}
\\
\Lambda \partial_{\Lambda} \tilde{g}_{\Lambda}
&=
\left(-1 +\frac{5}{4} \eta^A_{\Lambda} + \frac{1}{4} \eta^Z_{\Lambda} \right) \tilde{g}_{\Lambda}+\frac{1}{8\pi} \tilde{g}^3_{\Lambda} \frac{1}{\left(1+\tilde{m}^2_{\Lambda}\right)^2}\;.
\label{eq:g}
\end{align}
The anomalous dimensions for the frequency and momentum factors, respectively, are different
\begin{align}
\eta^Z_{\Lambda} = -\frac{1}{48\pi} \tilde{g}^2_{\Lambda} \frac{1}{\left(1+\tilde{m}^2_{\Lambda}\right)^2}
\label{eq:eta_Z}\\
\eta^A_{\Lambda} = \frac{5}{16\pi} \tilde{g}^2_{\Lambda} \frac{1}{\left(1+\tilde{m}^2_{\Lambda}\right)^2}\;.
\label{eq:eta_A}
\end{align}
%

\subsection{Interacting fixed point}

In the infrared $\Lambda\rightarrow 0$, these flow equations have two fixed points. A trivial, non-interacting fixed point 
$\tilde{g}_{\Lambda\rightarrow0}=0$ with zero 
anomalous dimensions. More interesting is the infrared-stable, 
interacting fixed point ($\tilde{g}_\ast \neq 0 $) with finite anomalous dimensions: 
\begin{align}
\eta^Z_\ast=-0.04
\;,\;\;\;\;\;\;
\eta^A_\ast=0.61\;.
\end{align}
We note that, as usual, the precise numerical values of the critical exponents depend on the renormalization scheme 
adopted, for example on the choice of the cutoff function. 

From the $\beta$-functions Eq.~(\ref{eq:mass}-\ref{eq:eta_A}), we see that the interacting fixed point separates a massive phase from a region of the phase diagram (not studied in the present work and denoted by a dashed line in Fig.~\ref{fig:phase_diag}) where the $\mathcal{P}\mathcal{T}$-symmetry is spontaneously broken. 

The existence of a non-Gaussian fixed point with finite $\tilde{g}^\ast$ is ensured by two conditions: (i) the relevant dimensional running leading to the $-1$ in the first bracket of Eq.~(\ref{eq:g}) and (ii) the positivity of the second term $\sim \tilde{g}^3$. In terms of power counting, (i) is actually similar to the conventional $\phi^4$ Wilson-Fisher fixed point in 3 dimensions 
\cite{chaikin_book}.
(ii) is a direct consequence of the imaginary nature of the cubic interaction in the Lagrangian Eq.~(\ref{eq:lattice}). The same mechanism turns out to stabilize the real-valuedness of the fixed points of the local $i\phi^3$ theory \cite{bender12}. Another signature of the non-unitary interaction is a negative anomalous dimension as it was also pointed out in reference \cite{bender12} and previously in the classic analysis of a $\phi^3$ theory by Fisher \cite{fisher78}. We also find a negative $\eta^Z$ for the frequency renormalization factor here in Eq.~(\ref{eq:eta_Z}).

A distinctive feature of the non-unitary quantum chain considered here is that it is not relativistically invariant nor it seem to become so at the critical point, at least not within our one-loop approximation. The cubic vertex depends linearly on spatial momenta (see below Eq.~(\ref{eq:lattice})), since it couples neighboring sites. By contrast, it does not depend in the same way on frequencies because the action is completely local in time. While such frequency-dependent interaction will also be generated in the RG flow, they appear only at higher-order in a loop expansion. As a consequence, at the one-loop order considered here, the anomalous dimension for the spatial momenta, $\eta^A_\ast$, is significantly larger and has the opposite sign to $\eta^Z_\ast$.

\medbreak

\section{Observable signatures}
\label{sec:observables}

At the critical point, the zig-zag phonon propagator takes the form:
\begin{align}
G^\ast_{\Lambda\rightarrow 0}(\omega,k) = \frac{1}{\omega^{2-\eta^Z_\ast} + k^{2-\eta^A_\ast}}\;,
\label{eq:phonon_prop}
\end{align}
resulting in a fractional value for the dynamical exponent at the interacting fixed point
\begin{align}
z_\ast=1+\eta^Z_\ast-\eta^A_\ast = 0.35\;.
\label{eq:z}
\end{align}
This strongly modifies the phonon dispersion at the critical point to 
\begin{align}
\omega_{\text{critical}} \sim k^{z_\ast} =  k^{0.35}
\end{align}
and leads to a critical continuum in contrast to the well-defined quasi-particle peak of the massive phase 
in Eq.~(\ref{eq:massive}). We observe that, upon approaching the critical point, when the correlation length 
in the chain becomes progressively larger, an exotic dispersion relation for the 
transverse vibrations appears that is neither a Luttinger liquid-type phonon, nor a massive 
short-ranged excitation, nor a long-ranged Ising spin-wave.

In trapped ion systems, several technologies are available to detect scaling 
properties associated with such 
quantum phase transitions \cite{islam11,ulm13}. 
The softening of the Zigzag mode can traced by Raman spectroscopy. In the 
phase with spontaneously broken $\mathcal{P}\mathcal{T}$-symmetry, the structural 
deformation of the chain in a Zigzag pattern can be measured via fluorescence 
measurement of the structure factor \cite{waki92,birkl92,shimshoni11,ulm13}.
Finally, it may be possible to detect the long-range entanglement in the 
quantum fragile regime by mapping the phonon dynamics back to spin 
states cite \cite{kim09}. 
We also note here that the opportunities of trapping ions with 
optical potentials \cite{schneider10,karpa13,linnet12,enderlein12} 
and usage of optical cavities \cite{cormick12}, may offer new routes toward engineering 
trapped ions systems with dissipative interactions.


\medbreak

\section{Conclusion and outlook}
\label{sec:conclu}

In this paper, we computed the quantum critical dynamics of transverse vibrations in a one-dimensional chain of trapped ions 
interacting via a cubic nonlinearity. The critical point we found may be viewed as a quantum fragile state of matter with unusual mechanical excitations. While we have no mathematical proof that the $\mathcal{P}\mathcal{T}$-symmetric $i\phi^3$ quantum chain has a well defined quantum mechanical ground state, it seems reasonable to extrapolate the work of Bender {\it et al.} \cite{bender12} on the local $i\phi^3$ theory to a one dimensional lattice with translationally-invariant nearest-neighbor interactions.

In the future, one could study local quantum quenches of Eq.~(\ref{eq:hamilton}) in the strongly nonlinear 
regime to investigate quantum analogue of classical solitary waves \cite{miri12, nesterenko84} and shocks \cite{gomez12}. These supersonic exications arise naturally in a fragile mechanical medium (with vanishing elastic moduli and {\it linear} speed of sounds) and they have a distinct origin from topological solitons generated by suitable choices of boundary conditions or sample topology \cite{landa10}. As we have demonstrated, quantum-fluctuations induce rigidity and generate non-linear 
excitations even in a mechanical system that was originally floppy. We expect strongly non-linear solitary waves to interact with such quantum background. We expect the amplitude of the solitary wave to be reduced as it propagates; eventually 
the soliton will decay.

We note that, while we here considered a cubic nonlinearity, 
we expect very interesting quantum fragile behavior to emerge for other nonlinearities as well. A quantum version 
of the Fermi-Pasta-Ulam problem \cite{gallavotti08}, for example, would be an interesting future problem. 
From an implementation viewpoint, it would be interesting to investigate how to generate power-law type 
interactions as those of the Hertz law, Eq.~(\ref{eq:hertz}) for overlapping soft spheres with ultracold atom systems. This would 
open the path to simulate the sonic vacuum in the quantum regime.

\begin{acknowledgments} 

We are grateful to R. Islam for explanations and references on trapped ion experiments.
We also thank S. Diehl, E. G. Dalla Torre, T. Giamarchi, M. Lukin, M. Punk, S. Sachdev, A. M. Turner, and 
P. Zoller for very helpful discussions. This research was supported by the DFG under grant Str 1176/1-1, by the NSF under Grant DMR-1103860, by the Army Research Office Award W911NF-12-1-0227, by the Center for Ultracold Atoms (CUA) and by the Multidisciplinary University Research Initiative (MURI).

\end{acknowledgments}

\end{document}